\documentstyle[12pt]{article}
\newcommand{\be}{\begin{equation}}
\newcommand{\ee}{\end{equation}}
\newcommand{\bea}{\begin{eqnarray}}
\newcommand{\eea}{\end{eqnarray}}
\newcommand{\bean}{\begin{eqnarray*}}
\newcommand{\eean}{\end{eqnarray*}}
\newcommand{\ba}{\begin{array}{l}}
\newcommand{\ea}{\end{array}}
\newcommand{\bb}{\begin{thebibliography}{99}}
\newcommand{\eb}{\end{thebibliography}}
\newcommand{\ci}[1]{\cite{#1}}

\newcommand{\Tr}{\mbox{Tr\,}}
\newcommand{\Ds}{\displaystyle}

\begin{document}
$\left.\right.$
\vskip 1cm
\Large
\begin{center}
{\bf On the higher-loop effective \\
action in NJL model }
\end{center}
\large
\centerline{ S.A.Garnov }
\normalsize
\centerline{ Tashkent State University, }
\centerline{    Tashkent 700095, Uzbekistan }
\vskip 1cm
The 3-loop effective action and effective potential in
Nambu - Jona-Lasinio model are calculated. The problem of
vanishing contributions in the higher orders is discussed.
The general form of such contributions is obtained.

\large
\section{Introduction}
The generating functional of 1PI Green's functions, which hereinafter
will be referred to as an effective action, is a convenient tool to
derive various dynamical information from the system we want to investigate.
Therefore it is important to have a method which makes it possible to
calculate the higher-loop effective action for models with both bosonic
and fermionic fields. Such a method based on the DeWitt formula \ci{dw} was
developed in recent works \ci{fm1} - \ci{fgg}, specifically,
in reference to bosonized
Nambu-Jona-Lasinio (NJL) model. The calculations were fulfilled up to
2-loop level. We want to continue using this method and, having made
some generalizations, apply it to calculations in higher orders.

The effective action obtained will be used to find the effective potential
for this model. Having calculated the effective potential, we can
investigate the spontaneous breaking of chiral symmetry
and the dynamical acquirement
of mass by fermionic field \ci{nj} - \ci{gn}.
The calculations carried out in \ci{fm2} displayed
an interesting property of NJL model, namely it turned out that 2-loop
contribution to the effective potential wanishes. We shall discuss this
problem in reference to higher orders and formulate and prove a theorem
about vanishing contributions. Also we shall make the direct calculations
of 3-loop corrections to the effective potential and show that there exsist
non-vanishing corrections among them.

The article is organised as follows. In Section 2 we describe the method
of calculations of the effective action. Section 3 is devoted to the
calculation of 3-loop effective potential. In Section 4 we obtain the
general form of vanishing contributions to the effective potential.

\section{Method of calculations}

In this section we shall review briefly the method based on the DeWitt
formula. In order to demonstrate how it works we shall take QED but
the results which will be obtained for QED are applicable after some
substitutions to NJL model. This fact will be widely used in the next
sections. We shall follow here the work \ci{fm2} and only at the end
of this section we make some generalisations which are important
for higher-loop calculations.

The generating functional for the spinor electrodynamics with Lagrangian

\be L = -\frac{1}{4} F_{\mu \nu} F^{\mu \nu} + \bar{\psi}
[ i(\gamma^{\mu} \partial_{\mu} - ie \gamma^{\mu} A_{\mu}) + m]\psi -
\frac{1}{2\alpha}(\partial_{\mu}A^{\mu})^{2} \ee
has the form (the source is introduced only for the photon field)
\be Z[J_{\mu}] = \int DA D\psi D\bar{\psi} exp [i\int L dx + i(JA)] \ee
Having integrated over spinor variables we get
\be Z[J_{\mu}] = \int DA  \exp [i S_{eff} + i(JA)] \ee
where
\be S_{eff} = S_{A} + i \hbar T \ee
\be \ba \Ds S_{A} = \frac{1}{2} A_{\mu} D^{-1}_{\mu \nu} A_{\nu} , \;
T = \Tr \ln K , \\
\Ds K^{-1} = i\gamma^{\mu} \partial_{\mu} + e \gamma^{\mu} A_{\mu}) - m,
\; D_{\mu \nu} = \frac{1}{\partial^{2}} (g_{\mu \nu} - (1 - \alpha)
\frac{\partial_{\mu} \partial_{\nu}}{\partial^{2}}) \ea \ee

Let us define the effective action in the following way
\be \Gamma[\langle  A_{\mu}\rangle] = W[J_{\mu}] - (J_{\nu}
\langle  A^{\nu}\rangle) \ee
where
\be \langle A_{\mu}\rangle =
Z^{-1} \int DA A_{\mu} \exp [i S_{eff} + i(JA)] \ee
is so-called "classical field" and
\be W[J_{\mu}] = -i \ln Z \ee
The effective action defined this way is the generating functional of
the Green's functions which are 1PI with respect to photon propagator.
From (3) and (6) we have
\be \frac{\delta \Gamma}{\delta \langle A_{\mu}\rangle} =
\langle \frac{\delta S_{eff}}{\delta A_{\mu}}\rangle \ee

For the models containing fermion and vector fields the general form of
DeWitt's formula is
\be \langle Q[A,\bar{\psi},\psi]\rangle = :\exp(G):
Q[\langle A\rangle,\langle \bar{\psi}\rangle,\langle \psi\rangle] \ee
where
\be \ba \Ds G = \frac{i}{\hbar} \sum_{n=2}^{\infty} \frac{(-i\hbar)^{n}}{n!}
\sum C_{ijk}^{n} (-1)^{j}
G^{\mu_{1}\ldots \mu_{i} \alpha_{1}\ldots\alpha_{j}\beta_{1}\ldots\beta_{k}}
\times \\ \Ds \times
\frac{\delta^{n}}{\delta\langle A_{\mu_{1}}\rangle\ldots
\delta\langle \psi_{\beta_{k}}\rangle
\ldots \delta \langle \bar{\psi}_{\alpha_{j}}\rangle\ldots} \ea \ee
\be
G^{\mu_{1}\ldots \mu_{i} \alpha_{1}\ldots\alpha_{j}\beta_{1}\ldots\beta_{k}}
= \frac{\delta^{n} W}
{\delta J_{\mu_{1}} \ldots \delta \eta_{\beta_{k}}
\ldots \delta \bar{\eta}_{\alpha_{j}} \ldots}  \ee
are the connected Green's functions with i photon and j+k fermion legs;
the colons mean that derivatives act only on $Q$; the Plank constant $\hbar$
is restored and
\be C_{ijk}^{n} = \frac{n!}{i! j! k!}, \; i+j+k=n \ee
Having applied (10) to the rhs of (9) we get
\be \frac{\delta \Gamma}{\delta \langle A_{\mu}\rangle} =  :\exp(G):
\frac{\delta S_{eff}[\langle A_{\mu}\rangle]}
{\delta \langle A_{\mu}\rangle} \ee
Below we shall deal only with $\langle A\rangle$ therefore
for convenience we shall
omit brackets keeping in mind that from
now on $A$ implies $\langle A\rangle$. In our
case we have for $G$
\be \ba \Ds G=
\frac{-i\hbar}{2}G^{\mu\nu}\frac{\delta^2}{\delta A^{\mu} \delta A^{\nu}} -
\frac{\hbar^{2}}{6}G^{\mu\nu\lambda}\frac{\delta^3}
{\delta A^{\mu} \delta A^{\nu} \delta A^{\lambda}} + \\
\Ds +\frac{i\hbar^{3}}{24}G^{\mu\nu\lambda\sigma}\frac{\delta^4}
{\delta A^{\mu} \delta A^{\nu} \delta A^{\lambda} \delta A^{\sigma}}
\ea \ee
where
\be
G^{\alpha_{1} \ldots  \alpha_{n}}
= \frac{\delta^{n} W}
{\delta J_{\alpha_{1}} \ldots \delta J_{\alpha_{n}}} \ee
It is convenient to use the following notation
\be
T^{\alpha_{1} \ldots  \alpha_{s}}
= (-1)^{s} \frac{\delta^{s} T}
{\delta A_{\alpha_{1}} \ldots \delta A_{\alpha_{s}}} \ee
Then
\be \frac{\delta S_{eff}}{\delta A_{\mu}} =
D^{-1\;\mu\nu} A_{\nu} - i\hbar T^{\mu} \ee
and we have from (14) up to the order of $\hbar^{3}$
\be \ba \Ds \frac{\delta \Gamma}{\delta A_{\mu}} =
D^{-1\;\mu\nu} A_{\nu} - \hbar \left[ iT^{\mu} +
\frac{\hbar}{2} G^{\nu\lambda} T_{\mu\nu\lambda} \right] - \\
\Ds \hbar^{2} \left[ \frac{i\hbar}{6} G^{\nu\lambda\sigma}
T_{\mu\nu\lambda\sigma} - \frac{i\hbar}{8} G^{\nu_{1}\lambda_{1}}
G^{\nu_{2}\lambda_{2}} T_{\mu\nu_{1}\lambda_{1}\nu_{2}\lambda_{2}}
\right] \ea \ee
We also need an equation to connect $G$ and $\Gamma$. This equation is
\be G^{\mu\nu}\frac{\delta^{2}\Gamma}
{\delta A^{\nu} \delta A^{\lambda}} = - \delta^{\mu}_{\lambda} \ee
Now we expand $G$ and $\Gamma$ in series over $\hbar$
\be \ba \Ds \Gamma = \Gamma_{0} + \hbar\Gamma_{1} +
\hbar^{2}\Gamma_{2} + \ldots \\
 \Ds G = G_{0} + \hbar G_{1} + \hbar^{2} G_{2} + \ldots \ea \ee
Here $\Gamma_{0}$ is the tree effective action, and $\Gamma_{1}$,
$\Gamma_{2}$, \ldots are the quantum corrections - 1-loop, 2-loop,\ldots
accordingly. Up to the 2-loop level the effective action was obtained in
\ci{fm2}:
\be \Gamma = \frac{1}{2} A_{\mu} D^{-1\;\mu\nu} A_{\nu} + i\hbar T +
\frac{1}{2}\hbar^{2} \Tr ( - D^{\mu\nu} T_{\mu\nu} ) \ee

Now we want to make some comments. The condensed notations used up to now
imply summarizing over discrete variables and integrating over continuos
ones in all the expressions. Nevertheless the question of how to calculate
traces over $\gamma$-matrices was out of discussion. Therefore the method
described above should be expanded with the appropriate rule in order to
satisfy the common rules of diagrammatic technique
(see for example \ci{ll}). Namely, the calculations
of traces must be carried out along every fermionic loop separately ( in our
case the loops are constructed not out of the simple propagators but out
of the propagators in "external field" ). If we have only one loop
this rule is satisfied by calculation of traces over the
whole expressions, but in the case of many loops it is not so.

\section{3-loop effective action}
Now we start investigating NJL model. We take this model in the
bosonized form
\be L = \bar{\psi} (i\gamma^{\mu} \partial_{\mu} + g(\sigma +
i\gamma_{5}\pi))\psi - \frac{1}{2} (\sigma^{2} + \pi^{2}) \ee
where $\sigma$ and $\pi$ are auxiliary scalar and pseudoscalar fields.
After integrating over spinor variables in the path integral we
have the following generating functional for this model
\be Z[J] = \int D\phi \exp (iS_{eff} + i(J\phi)) \ee
where we used the notations
\be \phi^{i} = \{ \sigma, \pi \},\; \gamma^{i} = \{ 1, i\gamma_{5} \},
\; \phi^{2} = \sigma^{2} + \pi^{2} \ee
and
\be \ba \Ds S_{eff} = -\frac{1}{2}\phi^{2} + i\hbar T, \\
\Ds T = \Tr \ln K, \; K^{-1} = i\gamma^{\mu} \partial_{\mu} + g\hat{\phi},
\; \hat{\phi} = \phi^{i}\gamma^{i} \ea \ee
Having introduced the free propagator of the auxiliary field $\phi^{i}$
as
\be D_{ij} (x-y) = \delta (x-y) \delta_{ij} \ee
we can re-write $S_{eff}$ as
\be S_{eff} = -\frac{1}{2}\phi^{i} D_{ij}^{-1}\phi^{j}+ i\hbar T \ee
It is easy to see that the NJL model re-writed in such a form becomes
analogous to QED after the substitutions
\be i,j,\ldots,\phi^{i} \rightarrow \mu,\nu,\ldots,A^{\mu} \ee
Therefore we get the effective action for NJL model (up to 2-loop level)
\be \tilde{\Gamma} = -\frac{1}{2}\phi^{i} D_{ij}^{-1}\phi^{j}+ i\hbar T
+ \frac{1}{2}\hbar^{2} \Tr ( - D^{i j} T_{i j} ) \ee
The corresponding diagrams are represented in Fig.1.
The auxiliary boson field is depicted by the dashed line, the thick
lines are the fermionic "propagators in external field".

Here we want to make some remarks about what we are calculating. In the
case of QED we implied according to (2) and (6) that there exist only
photon external fields and the fermion propagators occur only as
internal lines. After we turned to study of NJL model the effective
action obtained in such a manner is the generating functional of the
Green's functions with no free quarks in external lines. Of course,
these Green's functions will be irreducible with respect to the
propagators of the auxiliary field but actually we are not interested
in this fact because these propagators will reduce into the points
according to (27).

Let us calculate 3-loop effective action and effective potential.
In order to do it we return to Eq.(19) which was obtained for QED and
use the substitutions (29). We have
\be \frac{\delta \tilde{\Gamma}_{3}}{\delta \phi^{i}} =
-\frac{1}{2} G^{jk}_{1} T_{ijk} - \frac{i}{6} G^{jkl}_{0} T_{ijkl}
+\frac{i}{8} G^{j_{1}k_{1}}_{0} G^{j_{2}k_{2}}_{0}
T_{ij_{1}k_{1}j_{2}k_{2}} \ee
$G_{0}$ and $G_{1}$ can be found from the following equation
\be \left( G_{0}^{lj} + \hbar G_{1}^{lj} + \ldots \right)
\left( \frac{\delta^{2} \tilde{\Gamma}_{0}}
{\delta\phi^{j} \delta \phi^{k}}
+ \hbar \frac{\delta^{2} \tilde{\Gamma}_{1}}
{\delta\phi^{j} \delta \phi^{k}} +\ldots \right) = -\delta^{l}_{j} \ee
From (31) and (32) we get
\be \frac{\delta \tilde{\Gamma}_{3}}{\delta \phi^{i}} =
-\frac{i}{2} G^{jp}_{0}T_{pr} G^{rk}_{0}T_{ijk} -
+\frac{i}{8} G^{j_{1}k_{1}}_{0} G^{j_{2}k_{2}}_{0}
T_{ij_{1}k_{1}j_{2}k_{2}} \ee
To obtain the last equation we took into consideration that
\be G_{0}^{abc} =  G_{0}^{aj}G_{0}^{bk}G_{0}^{cl}
 \frac{\delta^{3} \tilde{\Gamma}_{0}}
{\delta \phi^{j} \delta \phi^{k} \delta \phi^{l}} = 0 \ee
on the strength of (30).

Therefore 3-loop contribution to the effective action for NJL model is
\be\ba \Ds \tilde{\Gamma}_{3} =
\frac{i}{4} G^{ik}_{0}T_{ij} G^{jl}_{0}T_{kl}
+\frac{i}{8} G^{ij}_{0} G^{kl}_{0} T_{ijkl} \\
\Ds G^{ij}_{0} = D^{ij} \ea \ee
The corresponding diagrams are shown in Fig.2.

Let us remind the comments made at the
end of Section 2. It is easy to see that in the left diagram in Fig.2
traces must be calculated over each fermionic loop separately
and this rule will not be
changed when we take into consideration that propagator $D^{ij}$ of
the auxiliary field assembles into a point according to (27) (Fig.3).
Therefore the contributions corresponding to left and right diagrams
in Fig.3 are not equivalent.

The effective potential can be obtained from the the effective action
by setting all the "classical fields" equal to constants. Thus we have
\be \tilde{\Gamma} = -\int dx V_{NJL} \ee

Let us write down separately the contributions corresponding to each
graphs in Fig.3 ($I_{(a)}$ represents left, and $I_{(b)}$ and $I_{(c)}$
accordingly central and right ones)
\be \ba \Ds I_{(a)} = g^{4} \int \prod_{i=1}^{4}
\frac{d^{d} p_{i}}{(2\pi)^{d}} \delta(\sum p_{i})
\Tr\left[ (\gamma^{\mu} p_{1\;\mu} + g\hat{\phi})^{-1} \gamma^{i}
(\gamma^{\mu} p_{2\;\mu} + g\hat{\phi})^{-1} \gamma^{j}\right] \times \\
\Ds \times
\Tr\left[ (\gamma^{\mu} p_{3\;\mu} + g\hat{\phi})^{-1} \gamma^{i}
(\gamma^{\mu} p_{4\;\mu} + g\hat{\phi})^{-1} \gamma^{j}\right]\ea \ee

\be \ba \Ds I_{(b)} = g^{4} \int \prod_{i=1}^{4}
\frac{d^{d} p_{i}}{(2\pi)^{d}} \delta(p_{2}+p_{3})
\Tr\left[ (\gamma^{\mu} p_{1\;\mu} + g\hat{\phi})^{-1} \gamma^{i}
(\gamma^{\mu} p_{2\;\mu} + g\hat{\phi})^{-1} \gamma^{i}\right. \times \\
\Ds \times
\left. (\gamma^{\mu} p_{3\;\mu} + g\hat{\phi})^{-1} \gamma^{j}
(\gamma^{\mu} p_{4\;\mu} + g\hat{\phi})^{-1} \gamma^{j}\right]\ea \ee

\be \ba \Ds I_{(c)} = g^{4} \int \prod_{i=1}^{4}
\frac{d^{d} p_{i}}{(2\pi)^{d}} \delta(\sum p_{i})
\Tr\left[ (\gamma^{\mu} p_{1\;\mu} + g\hat{\phi})^{-1} \gamma^{i}
(\gamma^{\mu} p_{2\;\mu} + g\hat{\phi})^{-1} \gamma^{j}\right. \times \\
\Ds \times
\left. (\gamma^{\mu} p_{3\;\mu} + g\hat{\phi})^{-1} \gamma^{i}
(\gamma^{\mu} p_{4\;\mu} + g\hat{\phi})^{-1} \gamma^{j}\right]\ea \ee
($d$ is the dimension of spice-time; we omit $\int dx$)

In order to calculate the traces we use the following representation
\be (\gamma^{\mu} p_{\mu} + g\hat{\phi})^{-1} =
(\gamma^{\mu} p_{\mu} -\frac{g}{p^{2}}
\gamma^{\mu} p_{\mu} \hat{\phi} \gamma^{\nu} p_{\nu})
 ( p^{2} - g^{2}\phi^{2})^{-1} \ee
which can be easily checked if we use
\be \hat{\phi} (\gamma^{\mu} p_{\mu}) \hat{\phi} (\gamma^{\nu} p_{\nu})
= \phi^{2} p^{2} \ee
After summarizing the traces there will be only the contributions
from $I_{(a)}$ and $I_{(c)}$ so we have the final form of 3-loop
effective potential
\be \ba \Ds V_{3} = - \frac{1}{2} d^{2} g^{4}
\int \left( \prod_{i=1}^{4} \frac{d^{d} p_{i}}{(2\pi)^{d}}
( p_{i}^{2} - g^{2}\phi^{2})^{-1} \right) (2\pi)^{d} \delta(\sum p_{i})
(p_{1}\cdot p_{2})(p_{3}\cdot p_{4}) \\
\Ds -\frac{1}{2} d^{2} g^{8} \phi^{4}
\int \left( \prod_{i=1}^{4} \frac{d^{d} p_{i}}{(2\pi)^{d}}
( p_{i}^{2} - g^{2}\phi^{2})^{-1} \right) (2\pi)^{d} \delta(\sum p_{i}) \\
\Ds -\frac{3}{2} d g^{6} \phi^{2}
\int \left( \prod_{i=1}^{4} \frac{d^{d} p_{i}}{(2\pi)^{d}}
( p_{i}^{2} - g^{2}\phi^{2})^{-1} \right) (2\pi)^{d} \delta(\sum p_{i})
[(p_{1}\cdot p_{2}) + (p_{3}\cdot p_{4})] \ea \ee

In order to make these integrals finite we should introduce a cut-off.
In the case of $d=2$ as the theory is renormalizable the cut-off
has no physical sense but if $d=4$ the cut-off becomes
a new phenomenological parameter of the thery. It may be fixed
by the various normalization conditions (see for example \ci{iim}).

The fact that $I_{(b)}$ is vanishing is the consequence of some general
properties of the model we have chosen. As a matter of fact, there exist a
species of vanishing contributions in the effective potential. In the
next section we shall formulate and prove the theorem which gives these
contributions in the most general form.

\section{Vanishing contributions}

\underline{\bf Theorem.}  {\sl The contributions to the effective potential
of NJL model given by the expressions like}
\be \ba \Ds I =  \int d^{d} p_{1} d^{d} p_{2} \{d^{d} q\}
 \delta(f({q})) \times \\
\Ds \times \Tr\left[\gamma^{i} (\gamma^{\mu} p_{1\;\mu}
+ g\hat{\phi})^{-1} \gamma^{i}
\ldots F(\{q\}) \ldots
(\gamma^{\mu} p_{2\;\mu} + g\hat{\phi})^{-1} \right] \ea \ee
{\sl are equal to zero}.

The graphic representations some of such contributions are shown in
Fig.4.
As is seen, the necessary conditions for diagrams to be vanishing are:
\newline
1) presence of at least 2 loops with independent momenta within them
(loops are constructed out of the "propagators in external field");
\newline
2) traces is calculated over the whole expression, not over any part
separately.

In order to prove it we use first the fact that momenta $p_{1}$ and
$p_{2}$ are not mixed with the other momenta $\{ q \}$ in the
$\delta$-function. Let us re-write (43) as
\be \ba \Ds I = \Tr\left[ \gamma^{i}
\left(\int d^{d} p_{1}
(\gamma^{\mu} p_{1\;\mu} + g\hat{\phi})^{-1} \right) \gamma^{i}
\right. \times \\
\Ds \times \left. \int \{d^{d} q\}  F(\{q\})
\left(\int d^{d} p_{2}
(\gamma^{\mu} p_{2\;\mu} + g\hat{\phi})^{-1} \right)
 \right]  \ea \ee
On the strength of (40) we get ($d$ is 2 or 4)
\be \ba \Ds
\int d^{d}p (\gamma^{\mu} p_{\mu} + g\hat{\phi})^{-1} =
\int d^{d}p (\gamma^{\mu} p_{\mu} -\frac{g}{p^{2}}
\gamma^{\mu} p_{\mu} \hat{\phi} \gamma^{\nu} p_{\nu})
 ( p^{2} - g^{2}\phi^{2})^{-1} = \\ \Ds =
\int d^{d}p ( -\frac{g}{p^{2}})
\gamma^{\mu} p_{\mu} \hat{\phi} \gamma^{\nu} p_{\nu}
 ( p^{2} - g^{2}\phi^{2})^{-1} \ea \ee
Further, as the cyclic permutations do not change the trace, we have
\be \ba \Ds I = \Tr\left[ \int \{d^{d} q\}  F(\{q\})
\int d^{d} p_{1} \int d^{d} p_{2} \right. \\
\Ds \left.
(\gamma^{\mu} p_{2\;\mu} \hat{\phi} \gamma^{\nu} p_{2\;\nu})
\gamma^{i}
(\gamma^{\mu} p_{1\;\mu} \hat{\phi} \gamma^{\nu} p_{1\;\nu})
\gamma^{i}
\frac{g^2}{p_{1}^{2}p_{2}^{2}}
 ( p_{2}^{2} - g^{2}\phi^{2})^{-1}
 ( p_{1}^{2} - g^{2}\phi^{2})^{-1}
 \right]  \ea \ee
Keeping in mind that $\gamma^{i} = \{ 1, i\gamma_{5} \}$ we get
\be \ba \Ds
(\gamma^{\mu} p_{2\;\mu} \hat{\phi} \gamma^{\nu} p_{2\;\nu})
\gamma^{i}
(\gamma^{\mu} p_{1\;\mu} \hat{\phi} \gamma^{\nu} p_{1\;\nu})
\gamma^{i} = (\sigma -i\gamma_{5}\pi) \gamma^{i}
(\sigma -i\gamma_{5}\pi) \gamma^{i} p_{1}^{2} p_{2}^{2} = \\
\Ds =(\sigma -i\gamma_{5}\pi)[(\sigma -i\gamma_{5}\pi) +
\sigma (i\gamma_{5})^{2} - i\gamma_{5}(i\gamma_{5})^{2}\pi]
p_{1}^{2} p_{2}^{2}=0 \ea\ee
Thereby we have proved that $I=0$. The proof is completed.

There is an obvious way to generalize the theorem. Namely,
{\sl the expressions like}
\be \tilde{I} =  \int d^{d} p_{1} d^{d} p_{2} \{d^{d} q\}
  \Tr \left[ \ldots \right]   \ldots \Tr \left[ \ldots \right] \ee
{\sl are equal to zero if at least one of the traces in the integrand
 can be reduced to the form given by the Eq.(42)}. An example
is shown in Fig.5.

The only graph which gives contribution to 2-loop effective
potential is the 2-loop diagram in Fig.4
(this graph can be derived from the right graph in Fig.1
after reducing the propagator of the auxiliary field).
Obviously, this is the very case
we discussed in the theorem above because the analitycal expression
for this graph has the form
\be
\int  \frac{d^{d} p_{1}}{(2\pi)^{d}} \frac{d^{d} p_{2}}{(2\pi)^{d}}
(\gamma^{\mu} p_{1\;\mu} + g\hat{\phi})^{-1}
(\gamma^{\mu} p_{2\;\mu} + g\hat{\phi})^{-1}  \ee
Therefore there exist no contributions
of the order $\hbar^{2}$ in the effective potential.
Another example is our Eq.(38) which determines contribution $I_{(b)}$
to 3-loop effective potential. On the other hand,
there exist other contributions in higher ($>2$) orders than those
represented in Fig.4, and as was shown by the direct calculations for
the order of $\hbar^{3}$ there are non-vanishing ones among them.

\section{Conclusion}

In this work we used the method based on the DeWitt formula to calculate
the effective action and the effective potential for bosonized NJL
model. Recently it was shown that 2-loop effective potential in this
model is equal to zero. As the value of effective potential is the
subject of importance when study phenomena like spontaneous breaking
of chiral symmetry it would be very attractive if the higher
contributions behaved themselves the same way, i.e. were vanishing too,
because in this case the (tree + 1-loop) effective potential would be
the precise result. But 3-loop calculations revealed non-vanishing
contributions which of course must be taken into consideration as
well as (if necessary) the higher (4-, 5-,...loop) ones.

The other interesting problem is which contributions are vanishing besides
2-loop one. The most general form of these contributions was obtained
(Eqs (43) and (48)) and discussed in the special theorem in Section 4.
As was shown, the crucial conditions for contributions to be vanishing
are the independence of the momenta within at least
2 loops and the manner of calculation of traces over $\gamma$-matrices.


\bb
\bibitem{dw} DeWitt B S 1965 {\sl Dynamical Theory of Groups and Fields}
(NY: Gordon and Breach).
\bibitem{fm1}  Faizullaev B A and Musakhanov M M 1993 {\sl
Turk.J.Phys.} {\bf 17} p 717.
\bibitem{fm2}  Faizullaev B A and Musakhanov M M 1995 {\sl
Ann.of Phys.} {\bf 241} p 394.
\bibitem{fgg}  Faizullaev B A, Galkin D V and Garnov S A 1996 {\sl
Talk at 10 Int. Conf. On Problems in QFT, Alushta, 13-18 May} .
\bibitem{nj} Nambu Y and Jona-Lasinio G 1961 {\sl Phys.Rev.}
{\bf 122} p 345.
\bibitem{gn} Gross D and Neveu A 1974 {\sl Phys.Rev.}
{\bf D10} p 3235.
\bibitem{ll} Landau and Lifshitz 1989 {\sl Course of Theoretical Physics
vol IV} (Moscow: Nauka).
\bibitem{iim} Iliopoulos J, Itzikson C and Martin A 1975 {\sl Rev.Mod.Phys.}
{\bf 47} p 165.
\eb

\newpage
\section*{Figures 1-3}
\newcounter{cms}
\setlength{\unitlength}{1mm}

\begin{picture}(158,55)
\thicklines
\put(20,25){\circle{30}}
\put(35,24){+}
\put(50,25){\circle{30}}
\thinlines
\multiput(43,25)(3,0){5}{\line(1,0){2}}
\thicklines
\put(65,25){,}
\put(70,25){\line(1,0){12}}
\put(85,24){=}
\thinlines
\put(90,25){\line(1,0){12}}
\put(105,24){+}
\put(110,25){\line(1,0){12}}
\multiput(115,25)(0,3){3}{\line(0,1){2}}
\put(125,25){\ldots}
\put(65,5){\bf Fig.1}
\end{picture}


\begin{picture}(158,60)
\thicklines
\put(20,35){\circle{30}}
\put(20,15){\circle{30}}
\thinlines
\multiput(13,17)(0,3){6}{\line(0,1){2}}
\multiput(27,17)(0,3){6}{\line(0,1){2}}
\thicklines
\put(50,25){\circle{30}}
\thinlines
\multiput(43,28)(3,0){5}{\line(1,0){2}}
\multiput(43,22)(3,0){5}{\line(1,0){2}}
\thicklines
\put(80,25){\circle{30}}
\thinlines
\multiput(73,25)(3,0){5}{\line(1,0){2}}
\multiput(80,18)(0,3){5}{\line(0,0){2}}
\put(65,5){\bf Fig.2}
\end{picture}


\begin{picture}(158,65)
\thicklines
\put(20,25){\circle{30}}
\put(30,25){\circle{30}}
\put(50,25){\circle{30}}
\put(65,25){\circle{30}}
\put(80,25){\circle{30}}
\put(100,25){\circle{30}}
\put(110,25){\circle{30}}
\thinlines

\put(10,35){\line(1,0){18}}
\put(17,36){\bf Tr}
\put(10,33){\line(0,1){2}}
\put(28,33){\line(0,1){2}}

\put(22,15){\line(1,0){18}}
\put(30,11){\bf Tr}
\put(22,15){\line(0,1){2}}
\put(40,15){\line(0,1){2}}

\put(44,15){\line(1,0){42}}
\put(63,11){\bf Tr}
\put(44,15){\line(0,1){2}}
\put(86,15){\line(0,1){2}}

\put(93,15){\line(1,0){24}}
\put(102,11){\bf Tr}
\put(93,15){\line(0,1){2}}
\put(117,15){\line(0,1){2}}

\thicklines
\put(40,6){\line(1,0){12}}
\thinlines
\put(54,5){=}
\thinlines
\put(58,6){\line(1,0){12}}
\put(73,5){+}
\put(78,6){\line(1,0){12}}
\put(83,6){$\vee$}
\put(92,5){\ldots}
\put(65,0){\bf Fig.3}

\end{picture}

\newpage
\section*{Figures 4-5}
\begin{picture}(158,80)
\thicklines
\put(20,45){\circle{30}}
\put(35,45){\circle{30}}
\put(60,45){\circle{30}}
\put(75,45){\circle{30}}
\put(90,45){\circle{30}}
\put(108,45){\ldots}

\thinlines

\put(11,55){\line(1,0){30}}
\put(21,56){\bf Tr}
\put(11,53){\line(0,1){2}}
\put(41,53){\line(0,1){2}}

\put(52,55){\line(1,0){45}}
\put(74,56){\bf Tr}
\put(52,53){\line(0,1){2}}
\put(97,53){\line(0,1){2}}

\thicklines
\put(60,20){\circle{30}}
\put(75,20){\circle{30}}
\put(85,20){\circle{30}}
\put(100,20){\circle{30}}
\put(118,20){\ldots}

\thinlines

\put(52,30){\line(1,0){55}}
\put(79,31){\bf Tr}
\put(52,28){\line(0,1){2}}
\put(107,28){\line(0,1){2}}

\put(65,0){\bf Fig.4}

\end{picture}


\begin{picture}(158,70)
\thicklines
\put(40,35){\circle{30}}
\put(40,15){\circle{30}}
\put(50,25){\circle{30}}
\put(60,25){\circle{30}}
\thinlines
\put(31,5){\line(0,1){40}}
\put(31,5){\line(1,0){2}}
\put(31,45){\line(1,0){2}}
\put(25,22){\bf Tr}
\put(52,15){\line(1,0){18}}
\put(52,15){\line(0,1){2}}
\put(70,15){\line(0,1){2}}
\put(58,10){\bf Tr}
\put(65,0){\bf Fig.5}
\end{picture}


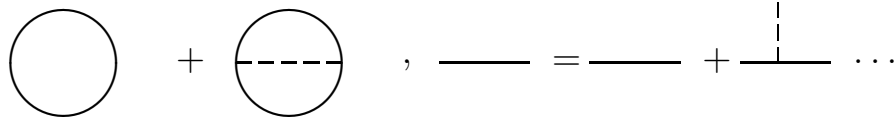
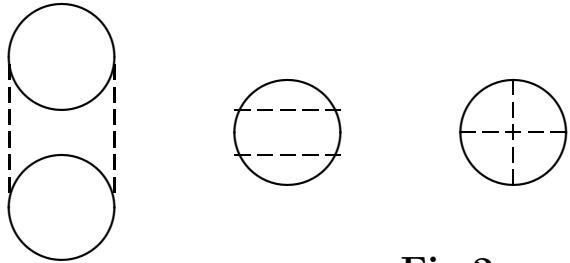
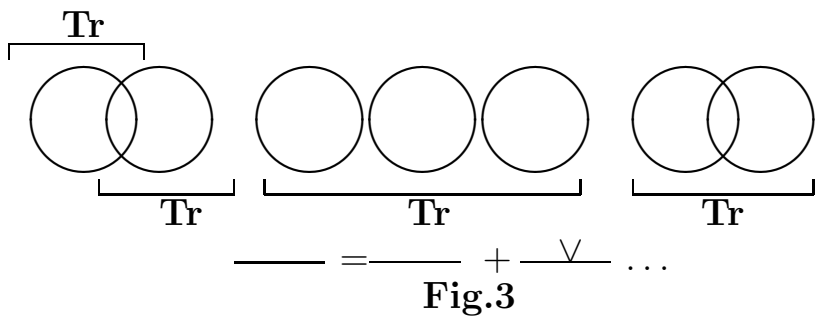
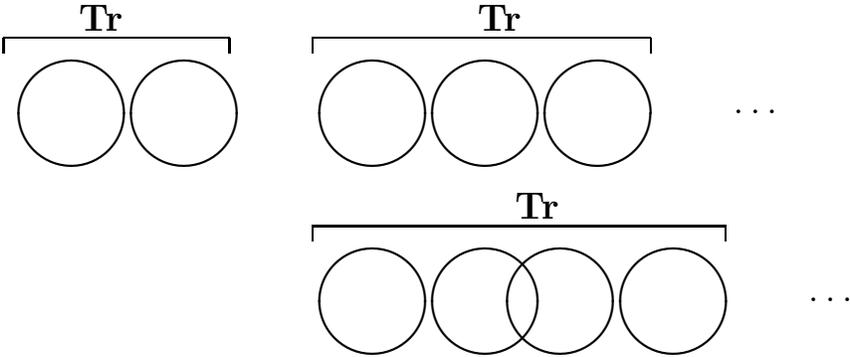
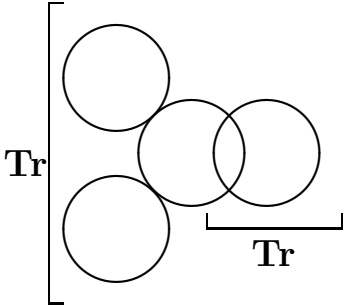
\end{document}